\documentclass[12pt]{JHEP3}

\let\ifpdf\relax
\usepackage{ifpdf}
\usepackage{graphics}
\usepackage{graphicx}
\usepackage{amssymb}
\usepackage{amsmath}
\usepackage{slashed}
\usepackage{cite}
\usepackage{epsfig}
\usepackage{datetime}

\newcommand{\be}{\begin{equation}}
\newcommand{\ee}{\end{equation}}
\newcommand{\bal}{\begin{align}}
\newcommand{\eal}{\end{align}}
\newcommand{\bea}{\begin{eqnarray}}
\newcommand{\eea}{\end{eqnarray}}

\def\p{\partial}
\def\a{\alpha}
\def\b{\beta}
\def\d{\delta}

\def\t{\theta}

\def\o{\omega}
\def\tphi{\tilde{\varphi}}
\def\O{\mathcal{O}}

\def\bz{\bar{z}}
\def\bp{\bar{\partial}}

\def\p{\partial}
\def\Tr{{\rm Tr}}

\def\tphi{{\tilde\varphi}}

\def\tphi{{\tilde \varphi}}

\def\tT{\tilde T}

\def\tr{{\rm tr}}
\def\eik{e^{2\pi i k/n}}
\def\ckn{C_{k/n}}
\def\bkn{\beta_{k/n}}
\def\skn{S_{k/n}}
\def\2kn{\frac{2\pi k}{n}}
\def\TiO{\Theta(0|i\Omega)}

\def\ToG{\Theta(0|\Gamma)}
\def\wllk{W_{1}^{1(k)}}
\def\wlzk{W_{1}^{2(k)}}
\def\wzlk{W_{2}^{1(k)}}
\def\wzzk{W_{2}^{2(k)}}

\def\suN{{{\rm SU}(N)}}
\def\sun1{{\widehat{\rm SU}(N)_1}}

\title{R\'enyi Entropy for the $\sun1$ WZW model on the torus}

\author{Howard J. Schnitzer  \\
{\small  Martin Fisher School of Physics, Brandeis University, \\ \ \ \ \ \ Waltham, MA 02454, USA\\
}

E-mail:  \email{schnitzr@brandeis.edu}}

\preprint{
 BRX-TH-6295}

\abstract{

The $\sun1$ WZW model is constructed on a n-sheeted branched torus, which allows the investigation of the R\'enyi entropy for a single interval at finite temperature. The small and large interval limits, as well as the low temperature expansion are presented for this theory.
}

\begin{document}

\section{Introduction}

Entanglement entropy  is an important tool in understanding various aspects of gravitational and quantum field theories. The entanglement entropy of a subsytem A is defined by the von Neumann entropy of the reduced density matrix
\bea
\label{EE}
S_{EE}=-\Tr \rho_A \log \rho_A
\eea
At finite temperature (\ref{EE}) has been studied for a variety of systems, which is a focus of this paper. In quantum field theory it is convenient to compute the R\'enyi entropy obtained by means of the replica trick, where the R\'enyi entropy is 
\bea
\label{1.2}
 S_n=-\frac{1}{n-1}\log\Tr \rho_A^{n}
\eea
and is related to the entanglement entropy by 
\bea
 S_{EE}=\lim_{n\to 1} S_{n} 
\eea
However such an analytic continuation may not be readily available, which is the case for the $\sun1$ WZW model. Therefore we are restricted to computing the R\'enyi entropy.

In two dimensional space-time one can consider the submanifold as a single interval, which we do. The two-dimensional Euclidean field theory is defined on a complex plane, so that the R\'enyi entropy becomes
\bea
\label{1.4}
S_n=-\frac{1}{n-1}\log (\frac{Z_n}{Z_1^n})
\eea
where $Z_n$ is the partition function for the n-sheeted Riemann surface which results from connecting the n-complex planes along the branch cuts, 
i.e. the interval. For a single interval at finite temperature the entanglement and R\'enyi entropies depend on the details of the CFT.
We are particularly interested in  $\sun1$ WZW theory, as it will allow future consideration of two important issues 1) the bose-fermi equivalence for the R\'enyi entropy and 2) the appropriate topological holographic dual. Both of these are under active consideration.

However $\sun1$ is interesting in its own right, as distinct from other 2d CFT of free bosons, as the   WZW models provide results which present a challenge for holography. Certainly, the usual $\sun1$ WZW theory is dual to a Chern-Simons theory. Is this true for the R\'enyi entropy for one or more intervals? 

The bose-fermi equivalence of the R\'enyi entropy for $\sun1$ raises the issue of the appropriate sum over characters in the fermi presentation, so as to coincide with that of the bose presentation. In this paper we focus on the bose construction of the R\'enyi entropy for $\sun1$ and discuss the fermi construction in future work.

There has been considerable work presenting material relevant and useful for this paper \cite{Chen:2015cna,Atick:1987kd,Lokhande:2015zma,Calabrese:2009ez,Calabrese:2010he,Chen:2014ehg,Cardy:2014jwa,Chen:2014hta,Datta:2013hba,Azeyanagi:2007bj,Chen:2013kpa,Chen:2015uia,Chen:2015usa,Caputa:2015tua,Chen:2015kua,Chen:2013dxa,Headrick:2010zt,Headrick:2012fk,Calabrese:2004eu,Calabrese:2009qy,Redlich:1986rp,Schnitzer:1986fi,Hartman:2013mia,Cardy:2007mb,DeNobili:2015dla,Casini:2009sr,Casini:2008wt,Casini:2012ei,Liu:2015iia}. The bosonic formulation of $\sun1$ is based on secs 4 and 5 of Naculich and Schnitzer \cite{Naculich:1989ii}. See also \cite{Redlich:1986rp,Schnitzer:1986fi}.  We are particularly indebted to refs \cite{Chen:2015cna,Atick:1987kd,Lokhande:2015zma,Calabrese:2009ez,Calabrese:2010he,Datta:2013hba} where we  often quote needed results from those papers, without repeating their derivations, in the interest of efficient discussion.

In Sec. \ref{2}, we present the bosonic action for $\sun1$ based on \cite{Naculich:1989ii}. A number of necessary group theoretic definitions and conventions are to be found in that section. Section \ref{3} presents twist operators appropriate to $\sun1$. In section 4, we present the classical and quantum solutions for the n-sheeted torus, resulting in the partition function $Z_n=Z_{n,\textrm{classical}}Z_{n,\textrm{quantum}}$. Since $Z_{n,\textrm{classical}}$ depends on the square of Riemann-Siegel theta functions $|\Theta(0|i\Omega)|^2$, continuation to $n\to1$ was not possible, so that we are unable to obtain the entanglement entropy. .

In Sec. \ref{5}, we present the small interval and large interval limits and in Sec. \ref{6} the low-temperature expansion of the theory. A summary of  relevant issues is found in Sec. \ref{7}.

\section{The Action \label{2}}
We begin with the action for $\sun1$ and  toroidal compactification, following Sec. \ref{5} of ref \cite{Naculich:1989ii}. Consider a free bosonic field $\varphi^{\mu}$ valued on a d-dimensional torus $T^d=\mathbb{R}^d/2\pi \Lambda$, with the torus obtained by identifying points in $\mathbb{R}^d$, differing by a point on the lattice $2\pi \Lambda$, where $\Lambda \subset \mathbb{R}^d$, generated by basis vectors $e_i$, $i=1$ to $d$. The dual vectors $e^{*i}$ are defined by $e^{\mu}_ie_{\mu}^{*j}=\delta_i^j$, where the metric on $\mathbb{R}^d$ is $\delta_{\mu \nu}$, $\mu,\nu=1$ to $d$. The bosonic field satisfies $\varphi^{\mu}\sim \varphi^\mu+2\pi n^ie_i^{\mu}$, for any set of integers $n^i$. The action for this field is 
\bea
\label{21}
S[\varphi]=\frac{1}{2\pi}\int (2i dz\wedge d\bz )(\p \varphi^{\mu}\bp \varphi^{\mu}+B_{\mu \nu}\p \varphi^{\mu}\bp \varphi^{\nu})
\eea
where $B_{\mu \nu}$ is a constant anti-symmetric tensor field. Define the metric $g_{ij}=e^\mu_ie^\mu_j$, with inverse $g^{ij}=e^{*i}_\mu e^{*j}_\mu$, so that the action becomes 
\bea
S[\varphi]=\frac{1}{2\pi}\int (2i dz\wedge d\bz )(g_{ij}+b_{ij})\p \varphi^i\bp \varphi^j
\eea
where $\varphi^\mu=\varphi^i e_i^\mu$ and $B_{\mu\nu}=b_{ij}e^{*i}_\mu e^{*j}_\nu$

Let $\Lambda_R$ be the root lattice of $\suN$, with simple roots $\a_i^\mu$ normalized such that the Cartan matrix is
\begin{equation}
M_{ij}=\a_i\cdot \a_j=\begin{cases}
 2&  {\rm for} \quad i=j \\
 -1 \quad & {\rm for} \quad |i-j|=1 \\
 0 \quad & {\rm for} \quad {\rm otherwise} 
\end{cases}\label{23}
\end{equation}
The weight lattice $\Lambda_W$ is generated by the weight vectors $\omega_\mu^i$ dual to $\a_i^\mu$ such that 
\bea
\o_i\cdot \a_j &=&\o_i^\mu \a_j^\mu=\d_{ij} \qquad {\textrm{and}} \nonumber  \\
\o_i\cdot \o_j &=& {\textrm{min}(i,j)-\frac{ij}{N}} 
\eea
The action (\ref{21}) is equivalent to that of $\sun1$ if the lattice $\Lambda_W$ is one-half the root lattice of $\suN$, so that $e_i^\mu=\frac12 \a_i^\mu$, $e_\mu^{*i}=2\o_\mu^i$, $g_{ij}=\frac 14 M_{ij}$, and 
\begin{equation}
b_{ij}=\begin{cases}
 \frac14 M_{ij} \quad & {\rm if} \quad i<j \\
 0 &  {\rm if} \quad i=j \\
 -\frac14 M_{ij} \quad & {\rm if} \quad i>j 
\end{cases} 
\end{equation}
The central charge of the theory is $c=N-1$, and $\tr M=2 c$. Each element of a representation $R_{I}$ of $\suN$ is associated with a weight vector $\o_\mu^I\in \Lambda_W$. In fact, the only allowed representations of $\sun1$ are the N unitary (integrable) representations $\hat{ R}_r$,  $r=0,\cdots , N-1$ which transform as the $r^{\rm{th}}$ fundamental representation of $\suN$ whose Young tableau is a single column of $r$ boxes.

In this paper we consider the single interval R\'enyi entropy  on a two-dimensional torus for $\sun1$. This torus will allow the consideration of R\'enyi entropy at finite temperature, and should not be confused with the torus of the target space of $\varphi^\mu$. One adopts the replica trick to compute $\tr \rho_A^n$ for integer n, where $\tr \rho_A^n$ is the partition function on an n-sheeted Riemann surface obtained by joining sucessive n-sheets along the region $A$. For a single interval we denote the surface as ${ \cal{R}}_{n,1}$. Thus, we are considering an n-sheeted branched torus.

The Lagrangian density does not depend explicitly on the Riemann surface, so that the structure of  ${ \cal{R}}_{n,1}$ is implemented by appropriate boundary conditions. The partition function is 
\bea
\label{26}
Z_{ { \cal{R}}_{n,1}}=\int[d\varphi] \exp\{-S_n[\varphi]\}
\eea
where $S_n[\varphi]$ is obtained from the Lagrangian density, now evaluated on the Riemann surface  ${ \cal{R}}_{n,1}$.

Following ref\cite{Calabrese:2009ez,Calabrese:2010he}, we consider n-independent copies of $\sun1$ and the partition function (\ref{26}) rewritten as a path integral on the n-sheeted torus
\bea
\label{27}
Z_{ { \cal{R}}_{n,1}}=\int_{{\mathbb{C}}_1}[d\varphi_1\cdots  d\varphi_n] \exp\left\{-\frac1{2\pi} \int_{{\mathbb{C}}} (2i dz\wedge d\bz )[{\cal L}[\varphi_1] +\cdots +{\cal L}[\varphi_n]   ]\right \}
\eea
where $\int_{{\mathbb{C}}_1}$ is the restricted path integral with boundary conditions
\bea
\varphi_i(x,0^+)=\varphi_{i+1}(x,0^-)\,,
\eea
where one identifies $n+1=1$.

\section{Twist Operators \label{3}}
Twist fields enforce two opposite permutation symmetries
\bea
i\to i+1, \quad \rm{and}\quad i+1\to i \quad (i=1, \cdots n)\,, \nonumber
\eea
identifying $n+1\equiv 1$. Thus
\bea
\label{3.1}
T_n=i\to i+1 \quad {\rm mod}\quad  n \nonumber  \\
\tilde{T}_n=i+1\to i \quad {\rm mod}\quad  n 
\eea
where $\tilde{T}_n$ can be identified with ${T}_{-n}$. 

Thus for the $j^{\rm{th}}$ sheet, circling the branch point $(z,\bz)$, one has 
\bea
\label{3.2}
\varphi_j^{\mu}(ze^{2\pi i},\bz e^{-2\pi i})=\varphi_{j+1}^{\mu}(z,\bz)
\eea
 which is implemented by $T_n$.
 
It is useful to introduce the linear combination 
 \bea
 \label{3.3}
 \tphi^{\mu}_k
=\sum_{j=0}^{n-1}\left(e^{2\pi i k/n }\right)^j \varphi^{\mu}_j 
\eea
which gets multiplied by $e^{2\pi i k/n }$ circling the twist operator. Thus (\ref{3.3}), diagonalizes the twist
\bea
\label{3.4}
T_n \tphi^{\mu}_k&=&e^{2\pi i k/n} \tphi_k^{\mu} \nonumber \\
\tilde{T}_n \tphi^{\mu}_k&=&e^{-2\pi i k/n} \tphi_k^{\mu}
\eea
 One can write
 \bea
 T_n &=&\prod_{k=0}^{n-1} T_{n,k} \qquad \rm{and} \nonumber \\
 \tT_n &=&\prod_{k=0}^{n-1} \tT_{n,k} 
 \eea
 with the  action of $T_{n,k}$  diagonal. For a single scalar field
\bea
T_{n,k} \tphi_{k'} &=&\tphi_{k'} \qquad {\rm if}\qquad k\neq k' \qquad \rm{and}\nonumber \\
T_{n,k} \tphi_k &=& e^{2\pi i k/n} \tphi_k
\eea
 Define 
 \bea
 \theta_{k}&=&e^{2\pi i k/n},\qquad \quad \textrm{so that} \nonumber \\
 \sum_{j=0}^{n-1}(\t_k)^j&=&0
 \eea
 For $\sun1$ the scalar fields are valued on the lattice $\Lambda_W$, so that one can write 
 \bea
 \tphi^{\mu}_k(z,\bz)=\tphi^a_k(z,\bz)e^\mu_a \qquad \qquad a=1 \,\, {\rm to} \,\, N-1
\eea
 
 It is useful to further diagonalize the twist operator with respect to the `` color''. That is 
 \bea
 T_{n,k} &=&\prod_{i=1}^{N-1} T^{(i)}_{n,k} 
 \eea
 where 
 \bea
 \label{3.11}
 T^{(i)}_{n,k}=\exp[i\zeta_\mu^{(i)}(k)\tphi^{\mu}_k(z)] 
 \eea
 with 
 \bea
 \label{3.12}
 \zeta_\mu^{(i)}(k)=\frac 12 ({k}/{n})\a_\mu^i ;\qquad \quad \mu=1 \,\, {\rm to }\,\, N -1
 \eea
 The two-point function satisfies 
 \bea
 \lim_{z\to z'}\langle \tphi_k^\mu(z)\tphi_l^\nu(z')  \rangle=-\delta_{kl} g^{\mu \nu} \ln|z-z'|\qquad \quad 0\le k,l \le n-1
 \eea
 Then the quantum dimension of $T_{n,k}$ is obtained from 
 \bea
 \label{3.14}
 \lim_{z\to z'}\prod_{i=1}^{N-1}\langle T_{n,k}^{(i)}(z)\tT_{n,k}^{(i)}(z')  \rangle &=&\exp-\left[\sum_{i=1}^{N-1} \frac 14 (k/n)^2 M_{ii}\ln|z-z'|\right] \nonumber \\
 &=&|z-z'|^{-\Delta_{k/n}}
 \eea
\bea
\label{3.15}
{\rm with} \qquad \Delta_{ k/n}=\frac c2 (k/n)^2
\eea 
where $\tr M=2c$ has been used, and $c=N-1$. 
 
 The scalar fields can be divided into a classical and quantum part, $\hat{\phi}_k^\mu$ and $\phi_k^\mu$ respectively, where 
 \bea
 \label{3.16}
 \tphi_k^\mu(z,\bz)=\hat{\phi}_k^\mu(z,\bz)+\phi_k^\mu (z,\bz)
 \eea
 where the quantum part is independent of the lattice contribution in computing the monodromies of the fields around a branch point. We examine these two contributions separately in the next section.
 
 \section{The partition function \label{4}}
 
 \subsection{The classical solution}
 From (\ref{3.16}),  the classical solution satisfies 
 \bea
\phi_k^{\mu}(ze^{2\pi i},\bz e^{-2\pi i})= \t_k \phi_{k}^{\mu}(z,\bz)+v_{(k)}^\mu 
\eea
 where $v_{(k)}^{\mu} \in \Lambda_{k/n}^\mu$, with the lattice defined by
 \bea
 \Lambda_{k/n}^\mu= \left\{q^\mu= \sum_{j=0}^{n-1}\t_k^j n_j^\mu \right\}
 \eea
 where $\mu=1 \,\, { \rm to}\, \,  N-1$ and $n_j^\mu \in \mathbb{Z}$. This generalizes eqns. (30) and (31) of \cite{Calabrese:2009ez}. The monodromy conditions satisfied by the classical solution on the world sheet are
 \bea
\oint _{\gamma_a}dz \p \phi_k^\mu  +\oint _{\gamma_a}d\bz \bp \phi_k^\mu  =v_{(k),a}^{\mu}  \\
\oint _{\gamma_a}dz \p \bar{\phi}_k^\mu  +\oint _{\gamma_a}d\bz \bp \bar{\phi}_k^\mu  =\bar{v}_{(k),a}^{\mu}  
 \eea
 where $\bar{\phi}_k^\mu$ is the complex conjugate of $\phi^\mu_k$, and $\gamma_a$ are the two cycles of the world-sheet torus with  $a=1,2$.
 
 The classical solution can be expressed in terms of the cut-differentials [see Appendix A] \cite{Chen:2015cna,Atick:1987kd}
 \bea
 \label{4.5}
\p \phi_k^\mu&=&a_{(k)}^\mu \,\o_1^{(k)}(z)  \nonumber \\
\bp \phi_k^\mu&=&b_{(k)}^\mu \, \bar{ \o}_2^{(k)}(\bz)  \nonumber \\
\p \bar{\phi}_k^\mu&=&\bar{a}_{(k)}^\mu \, \o_2^{(k)}(z)  \nonumber \\
\bp \bar{\phi}_k^\mu&=&\bar{b}_{(k)}^\mu \, \bar{\o}_1^{(k)}(\bz) \qquad \qquad \mu = 1 \quad {\rm to}\quad N-1
 \eea
 The solution to monodromy conditions are \cite{Chen:2015cna,Atick:1987kd,Lokhande:2015zma,Datta:2013hba}
 \bea
 \label{4.6}
 a_{(k)}^\mu&=&[W_2^{2(k)}\bar{v}_{(k),1}^{\mu}-W_1^{2(k)}v_{(k),2}^{\mu}]/{\rm det}W^{(k)} \nonumber \\
 \bar{a}_{(k)}^\mu&=&[\bar{W}_1^{1(k)}\bar{v}_{(k),2}^{\mu}-\bar{W}_1^{1(k)}\bar{v}_{(k),1}^{\mu}]/{\rm det}\bar{W}^{(k)} \nonumber \\
 b_{(k)}^\mu&=&[-W_2^{1(k)}v_{(k),1}^{\mu}+W_1^{1(k)}v_{(k),2}^{\mu}]/{\rm det}W^{(k)} \nonumber \\
 \bar{b}_{(k)}^\mu&=&[\bar{W}_2^{2(k)}\bar{v}_{(k),1}^{\mu}-\bar{W}_1^{2(k)}\bar{v}_{(k),2}^{\mu}]/{\rm det}\bar{W}^{(k)}
 \eea
 
 Next we present the Lagrange density and classical action appropriate to (\ref{4.5}), and  (\ref{4.6}).
 The Lagrange density (\ref{27}) for the classical solution leads to the classical portion of the action at fixed $k$,
 \bea
 \label{4.7}
 S^{(k)}=\frac{4}{n \pi}\frac{W_1^{1(k)}W_2^{2(k)}}{[{\rm det}W^{(k)}][{\rm det}\bar{W}^{(k)}]}\sum_{a,b=1}^{N-1}(g_{ab}+b_{ab})[(v_{(k),1}^a \bar{v}_{(k),1}^b)|W_2^{2(k)}|^2+(v_{(k),2}^a \bar{v}_{(k),2}^b)|W_1^{1(k)}|^2] \nonumber \\
 \eea
 As emphasized by \cite{Chen:2015cna,Atick:1987kd}, once one fixes the monodromy around the canonical cycles, one can obtain all the monodromies on the Riemann surface. For the n-sheeted torus these can be fixed as 
 \bea
\label{4.8}
 \oint _{\gamma_1}dz \p \phi_j^\mu&=&2\pi m_j^\mu \qquad {\rm and} \nonumber \\
 \oint _{\gamma_2}dz \p \phi_j^\mu&=&2\pi l_j^\mu  
 \eea
 where $m_j^\mu$ and $l_j^\mu \, \, \in \,\, \mathbb{Z}$.
 In the basis in which the twist operators acts as in (\ref{3.4}) one has 
 \bea
 \label{4.9}
v_{(k),1}^a&= 2\pi&\sum_{j=0}^{n-1}(\eik)^j m_j^a \qquad {\rm and } \nonumber \\
v_{(k),2}^a&= 2\pi&\sum_{j=0}^{n-1}(\eik)^j n_j^a
 \eea
 The action becomes
 \bea
S^{(k)}=\frac{16\pi}{n  } \sum_{a,b=1}^{N-1}(g_{ab}+b_{ab}) \sum_{j=0}^{n-1}\sum_{j'=0}^{n-1}\{  \bkn \,(\eik)^{j-j'} 
\,(m_j^a m_{j'}^b) \nonumber \\ 
+(\bkn)^{-1}\,(\eik)^{j-j'}\,(n_j^a n_{j'}^b)  \}
 \eea
 where we have defined 
 \bea
 \label{4.11}
  \bkn=\left | \frac{\wzzk}{\wllk} \right |
 \eea
Also define
\bea
(\ckn)_{jj'}=\cos[\2kn(j-j')] \nonumber  \\
(\skn)_{jj'}=\sin[\2kn (j-j')]
\eea
The classical partition function is 
\bea
Z_n^{cl}=\exp -\sum_{k=0}^{n-1}S^{k}
\eea
The sum over $k$ in the total partition function at fixed $(j,j')$ involves two terms, $\skn$ and $(S)_{(n-k)/n}$, which cancel in the sum over $k$. Therefore effectively 
\bea
\label{4.14}
S^{(k)}=\frac{16\pi}{n} \sum_{j=0}^{n-1}\sum_{j'=0}^{n-1}\sum_{a=1}^{N-1}\{ \bkn\, m_j^a (\ckn)_{jj'}\,m_{j'}^a \nonumber \\ \qquad \qquad + (\bkn)^{-1}\, n_j^a (\ckn)_{jj'}\, n_{j'}^a\}
\eea
Carry out a Poisson re-summation of the second term in (\ref{4.14}), changing $n^a_j\to l_j^a$. Then 
\bea
\label{4.15}
Z_n^{(cl)}=\prod_{k=0}^{n-1}(\bkn)^{1/2}\sum_{m_j^a,l_j^a} \exp -\frac{16\pi}{n}\sum_{a=1}^{N-1}\sum_{j=0}^{n-1}\sum_{j'=0}^{n-1} (\bkn)\{ \vec{m}^a \cdot (\ckn)\cdot \vec{m}^a\nonumber \\
+  \,\,\vec{l}^a \cdot (\ckn)\cdot \vec{l}^a\}
\eea
where 
\bea
\vec{m}^a \cdot (\ckn)\cdot \vec{m}^a=m_j^a [g_{ab}(\ckn)_{jj'} ]m_{j'}^b
\eea
so that 
\bea
\qquad \vec{m}\,\, \in \, \, (\mathbb{Z}^n)^{N-1} 
\eea
 
 The Riemann-Siegel theta function is 
 \bea
 \label{4.18}
\ToG=\sum_{m\in \mathbb{Z}^{n}}\exp [i\pi m^\dagger \cdot \Gamma \cdot m]
 \eea
 In our case, from (\ref{4.15})
 \bea
 \label{4.19}
\Gamma_{ab,jj'}=\frac{16i}{n}\delta_{ab}\sum_{k=0}^{n-1}\bkn (\ckn)_{jj'}=i\Omega_{ab,jj'}
 \eea
Therefore, the classical partition function is 
\bea
\label{4.22}
Z_n^{cl}=\left[\prod_{k=0}^{n-1}\bkn\right]^{1/2}\left[\TiO\right]^2 
\eea

Recall  (\ref{4.18}) and (\ref{4.19}), and define
\bea
(p_L^\mu)_s=-m_s^a e_a^{\mu}
\eea
where $\mu$ and  $a=1 \,\, {\rm to} \,\, N-1$, and $s=0\, \, {\rm to}\, \, n-1$. Then write
 \bea
 \label{4.24}
(p_L^\mu)_s=p_s^\mu + \o_s^{\mu}
\eea
\bea
\label{4.25}&&{\rm where} \qquad p_s^\mu \, \, \in \,\, (\Lambda_R)^g \\
\label{4.26}&&{\rm and} \qquad \o_s^\mu \,\, \in \,\, (\Lambda^g_W/\Lambda_R^g)
\eea 
where  $\Lambda_R^g$ and $\Lambda_W^g$ are the root and weight lattices at fixed $s$, and $g=N-1$.  [Recall Sec. \ref{2}] Similarly define
\bea
(p_R^\mu)_s=l_s^a e_a^{\mu}
\eea
as well as the analogues of (\ref{4.25}) and (\ref{4.26}). Then 
\bea
\label{4.28}
Z_n^{cl}=\left[\prod_{k=0}^{n-1}\bkn\right]^{1/2}
\sum_{\vec{\o}_r} \left |\sum_{\vec{p}_R \in \Lambda_R^g} \exp - \pi \sum_{\mu, \nu =1}^{N-1} \sum_{r\,,s\,=0}^{n-1}(p_r +\o_r)^\mu \Omega_{\mu \nu , r,s} (p_s +\o_s)^\nu \right |^2 \nonumber \\
\eea
where $p_s^\mu$ is as in (\ref{4.24}), (\ref{4.25}) and $\o_r^\mu$ is as in (\ref{4.26}). [When $n=1$, this sum goes over holomorphic blocks. See \cite{Naculich:1989ii}, eqn (4.37)]. In particular in (\ref{4.28})
\bea
\vec{\o}_r=(\o_{r_1}, \cdots , (\o_r)_{N-1}) \,  \qquad r=0\, \,  {\rm to} \, \, n-1 
\eea
where each $\o_{r_i}$ is a fundamental weight corresponding to a $\suN$ Young 
tableau with a single column of boxes. 

\subsection{The quantum solution}
 
 The quantum portion of (\ref{3.16}) satisfies
 \bea
 \label{4.30}
\hat{\phi}^{\mu}_k(e^{2\pi i}z, e^{-2\pi i}\bz)=\t_k \hat{\phi}^{\mu}_k(z, \bz)
 \eea
 so that for any closed loop $C$.
 \bea
\Delta_c \hat{\phi}^{\mu}_k=\oint_c dz \p \hat{\phi}^{\mu}_k +\oint_c d\bz \bp \hat{\phi}^{\mu}_k =0
 \eea
 Following refs \cite{Chen:2015cna,Atick:1987kd,Lokhande:2015zma,Calabrese:2009ez,Calabrese:2010he,Datta:2013hba}   , the quantum contribution to the partition function is 
 \bea
 \label{4.32}
Z_{n, k}^{(qu)}=\frac{1}{|\eta(\tau)|^{2(N-1)n}}\prod_{k=0}^{n-1}\left\{ \frac{1}{|\wllk \wzzk|^{1/2}}\left[ \frac{\t_1'(0|\tau)}{\t_1(z_1-z_2|\tau)}\right]^{\Delta_{k/n}} \left[ \overline{\frac{\t_1'(0|\tau)}{\t_1(z_1-z_2|\tau)}}\right]^{\Delta_{k/n}}\right\} \nonumber \\
 \eea
 The short distance limit is
\bea
\left[ \frac{\t_1'(0|\tau)}{\t_1(l|\tau)}\right]^{\Delta_{k/n}}_{ }\longrightarrow l^{-\Delta_{k/n}} \quad \textrm{as} \quad{l \to 0}
\eea 
 which is consistent with (\ref{3.14})
 
 \subsection{ Partition function on the n-sheeted torus}
 
 Combining equation (\ref{4.22}) and (\ref{4.32}), the partition function on the n - sheeted torus is
 \bea
 \label{4.34}
Z_n&=&Z_n^{(cl)} Z_n^{(qu)} \nonumber \\
 &=& \frac{c_n}{|\eta(\tau)|^{2(N-1)n}}\prod_{k=0}^{n-1}\left\{ \frac{1}{|\wllk |}\left | \frac{\t_1'(0|\tau)}{\t_1(z_1-z_2|\tau)}\right |^{2\Delta_{k/n}} |\TiO|^2 \right\} 
 \eea
 with $c_n$ an overall normalization independent of $\tau$ and $(z_1 -z_2)$.
 This is the main result of this paper. 
 
 In the next section we isolate the vacuum module from (\ref{4.34}), which may be useful in examining holographic duals. Then in Secs. \ref{5} and \ref{6} we consider the small interval, large interval, and low-temperature limits of (\ref{4.34}), following the strategy of \cite{Chen:2015cna,Atick:1987kd}.
 
 \subsection{Vacuum module}
 
 It is interesting to isolate the vacuum module from (\ref{4.34}), obtained from the identity representation $\o_r^\mu=0$ for all $r$. One can express
 \bea
(p^\mu)_s=\sum_{a=1}^{N-1}(n_a)_s \,\a_a^\mu \,  \qquad s=0\, \,  {\rm to} \, \, n-1 
 \eea
 with $(n_a)_s \, \in \, \mathbb{Z}$. Then, using the properties of the Cartan matrix 
 \bea
 \label{4.36}
(p^\mu)_s (C_{k/n})_{s,t}(p^\mu)_t=2 \sum_{s,t=0}^{n-1}\left\{  \sum_{a=1}^{N-1} (n_a)_s (n_a)_t - \sum_{a=1}^{N-2} (n_{a+1})_s (n_a)_t\right\} (C_{k/n})_{st} \nonumber \\
 \eea
 One can then consider (\ref{4.34}) with (\ref{4.36}) in the large N limit to investigate a holomorphic interpretation of this sub-module \cite{Headrick:2015gba}.  This issue is outside the scope of this paper.

\section{Some limits \label{5}} 
\subsection{Small interval limit \label{5.1}}

We apply the strategy and equation (A-21) of \cite{Chen:2015cna,Atick:1987kd}, which gives the small interval limits 

\bea
\wllk &= &1+\O(z_1-z_2)\\
\wzzk &=& i\b +\O(z_1-z_2)
\eea
 Therefore
 \bea
\bkn\to \b
 \eea
 in this limit.
 
 Consider the partition function (\ref{4.34}) on a rectangular torus of spatial size $L$ and Euclidean time $\b$. The R\'enyi entropy (\ref{1.2}) is associated with a spatial region $A$ running from $0$ to $l$. From (\ref{4.15}) to (\ref{4.34}) with $c=N-1$,
 \bea
 \label{5.4}
Z_n=\frac{1}{|\eta (\tau)|^{2(N-1)n}} ({l}/{L})^{-\frac{c}{12}n(1-1/n^2)}\!\!\!\! \!\!\!\!\sum_{m_j, l_j \in (\mathbb{Z}^{N-1})^n}\!\!\! \exp -\pi [m\cdot \Omega \cdot m+l\cdot \Omega \cdot l]
 \eea
 Therefore\footnote{Don't confuse the small interval length $l$ with the vector $\vec{l} $ in (\ref{5.4})}
 \bea
\Omega_{ab,jj'}\to \frac{16 \,g_{ab} \, \b}{n} \,\sum_{k=0}^{n-1}(\ckn)_{jj'}
 \eea
 Further 
 \bea
\sum_{k=0}^{n-1}\left({\eik}\right)^{(j-j')}=n \delta_{jj'} \,,
 \eea
 so that after summing on $j$ and $j'$,
 \bea
Z_n \to \frac{1}{|\eta (\tau)|^{2(N-1)n}} ({l}/{L})^{-\frac{c}{12}n(1-1/n^2)}\!\! \left\{\sum_{m_a, l_a \in \mathbb{Z}^{N-1}}\!\!\! \exp -16\pi \b\, [\bar{m} \cdot\bar{ m}+\bar{l}\cdot  \bar{ l}]\right \}^n
 \eea
 Therefore
  \bea
Z_n \to ({l}/{L})^{-\frac{c}{12}n(1-1/n^2)}[(Z_1)^n+\O(l/L)]
 \eea
in agreement with \cite{Chen:2015cna,Atick:1987kd,Lokhande:2015zma} for other CFT's.
The entanglement entropy can be computed for this limit using (\ref{1.4}), since one can now carry out the analytic continuation in $n$.

\subsection{Large interval limit \label{5.2}}
One makes use of various modular transformations
\bea
\t_1(-k/n|i\b)&=&i \b^{-1/2}e^{-\frac{k^2}{n^2 \b}}\t_1\left(\frac{ik/n}{\b}\Bigg | \frac{i}{\b} \right ) \\
\eta(-1/\tau)&=&(-i\tau)^{1/2} \eta(\tau) \\
\frac{\t_1'(0|-1/\tau)}{\t_1(z/\tau|-1/\tau)}&=&i\tau e^{-i\pi z^2/\tau}\frac{\t_1'(0|\tau)}{\t_1(z|\tau)}
\eea
and 
\bea
\left |\frac{\wzzk}{\wllk}\right | \to \left |\frac{\wllk}{\wzzk}\right | 
\eea
See refs  \cite{Chen:2015cna,Lokhande:2015zma,Datta:2013hba}. Then following closely the arguments of \cite{Chen:2015cna,Lokhande:2015zma} applied to (\ref{4.34}), and  the torus of Sec. \ref{5.1}, one obtains 
\bea
Z_n \to ({l}/{L})^{-\frac{c}{12}n(1-1/n^2)}[Z_1(n\b)+\cdots]
 \eea
 Once again note that the entanglement entropy can be computed in this limit from (\ref{1.4}), as the $n$-dependence can be continued to $n \to 1$.
 
\subsection{Thermal entropy}
A straightforward calculation gives
\bea
\label{5.14}
\lim_{l/L \to 0} [S_{EE}(1-{l}/{L})-S_{EE}(l/L)]&=& -\lim_{n\to 1}\frac{1}{n-1}[\log Z_1(n\b)-n\log Z_1(\b)] \nonumber \\
&=& \log Z_1(\b)-\frac 1\b \frac{Z_1'(\b)}{Z_1(\b)} \nonumber \\
&=& S_{th}
\eea
where $S_{th}$ is the thermal entropy,
which is in agreement with other CFT's  \cite{Chen:2015cna,Lokhande:2015zma,Chen:2014ehg}, as well as the holographic discussion \cite{Azeyanagi:2007bj}.
 
 \section{Low temperature limit \label{6}}
 We begin with (\ref{4.18}), (\ref{4.19}) and (\ref{4.34}) and expand with respect to $q=e^{-2\pi \b}$, so that 
 \bea
\eta (\tau)=q^{1/24}(1+{\cal O}(q)) \\
\t'(0)=2\pi q^{1/8} (1+{\cal O}(q)) \\
\t_1(z_1-z_2|\tau)=2q^{1/8}\sin\pi(z_1-z_2)(1+{\cal O}(q))
 \eea
 From ref \cite{Chen:2015cna}\footnote{ The discussion of this section follows closely that of \cite{Chen:2015cna}, section 3.}, their equations (3.5) and (A.3),
 \bea
W_1^{1(k)}\to 1+{\cal O}(q)\,, \qquad \quad \rm{and}
 \eea
 \bea
\frac{W_2^{2(k)}}{W_1^{1(k)}}=i\b +\int_{z_1}^{z_2}dt \frac{2i [\sin \pi({ k}/{n})(t-z_1) ][\sin\pi(1-k/n)(t-z_1)]}{\sin\pi(t-z_1)} +{\cal O}(q)
 \eea
 The leading and next to leading contribution to (\ref{4.18}) and (\ref{4.19}) come from all $m_a=0$ and $m_a=\pm 1$ respectively, with 2$n$ choices for the later.
 Therefore 
 \bea
\Theta(0 | i\Omega)\approx  1+ 2n(N-1)e^{-\pi \b}\left[\frac{\sin \pi (z_2-z_1)}{n\sin \frac \pi n(z_2-z_1)}\right] +{\cal O}(e^{-2\pi \b})
 \eea
 Using this in conjunction with (\ref{4.34})
 \bea
Z_n \sim c_n \frac{1}{q^{n/6}}\left(\frac{\pi }{\sin \pi l/L}\right)^{\frac{c}{12}n( 1-1/n^2)}\left\{ 1+4n(N-1)e^{-\pi \b}\left[\frac{\sin \pi (l/L)}{n\sin \frac \pi n(l/L)}\right] \right\}
 \eea
 where $c_n$ is an overall normalization constant which does not depend on $q$ or $(z_1-z_2)$. Therefore, the R\'enyi entropy in this limit is 
 \bea
 \label{6.8}
S_n=\tilde{c}_n -\frac{1}{n-1}4n(N-1)\left(\left[\frac{\sin \pi (l/L)}{n\sin \frac \pi n(l/L)}\right] -1\right)e^{-\pi \b}+\cdots 
 \eea
where $\tilde{c}_n$ is an overall normalization, where (\ref{6.8}) is consistent with the universal behavior predicted by Cardy and Herzog \cite{Cardy:2014jwa}.

\section{Discussion \label{7}}
In this paper we considered the $\sun1$ WZW model on a n-sheeted branched torus, which allows the computation of the R\'enyi entropy for a single interval at finite temperature for the 2d CFT. The n-sheeted partition function $Z_n$ is given in terms of Riemann-Siegel functions, which  then does not allow an obvious analytic continuation for $n\to 1$, which thus prevents 
a computation of the entanglement entropy. However such an analytic continuation is possible for small and large interval limits, and the leading term of the low-temperature expansion.

The discussion of the $\sun1$ WZW model in this paper was given entirely in the bosonic formulation. It is known that there is a bose-fermi equivalence for the $\sun1$ WZW model for $Z_1$ on a genus $g$ surface \cite{Naculich:1989ii}. The fermi presentation of  $\sun1$ allows for discussion of the  R\'enyi entropy in the presence of a continuum constraint gauge field. This is presently under investigation.

It is known that for $n=1$, $\sun1$ is dual to a Chern-Simons topological theory. The same issue arises for the R\'enyi entropy of the theory and its dual interpretation. What is the detailed Chern-Simons dual relevant to the construction of this paper? See \cite{Gukov:2004id} for a discussion of a related issue.\footnote{We thank M. Headrick for bringing this paper to our attention.} In Sec. 4D we isolated the R\'enyi entropy of the vacuum subspace of $\sun1$. Does this by itself have a holographic interpretation? 

Another application of WZW models to problems of entanglement is \cite{Schnitzer:2015gpa} where left-right entanglement and level-rank duality is discussed. Bose-fermi duality and entanglement entropies for a massless Dirac fermion and a compact free boson in two dimensions is considered in  \cite{Headrick:2012fk}.

\appendix 
\section {W-functions}
The contents of this Appendix are from equation (2.9), (A.1) and (A.2) of \cite{Chen:2015cna},

Cut differentials are defined by (2.9) of \cite{Chen:2015cna}
\bea
\o_1^{(k)}(z)=\t_1(z-z_1|\tau)^{-(1-\frac kn)}\t_1(z-z_2|\tau)^{-\frac kn}\t_1(z-(1-\frac kn)z_1-\frac kn z_2)|\tau) \nonumber \\
\o_2^{(k)}(z)=\t_1(z-z_1|\tau)^{-\frac kn}\t_1(z-z_2|\tau)^{-(1-\frac kn)}\t_1(z-\frac kn z_1-(1-\frac kn)z_2)|\tau)
\eea

The W-functions are given by  equations (A.1) and (A.2) of \cite{Chen:2015cna}
\bea
\wllk=\int_0^1 dz \, \, \t_1(z-z_1|\tau)^{-(1-\frac kn)}\t_1(z-z_2|\tau)^{-\frac kn}\t_1(z-(1-\frac kn)z_1-\frac kn z_2)|\tau) \nonumber \\
\wlzk=\int_0^1 d \bz   \, \,  \bar{\t}_1(\bz-\bz_1|\tau)^{-\frac kn}\bar{\t}_1(\bz-\bz_2|\tau)^{-(1-\frac kn)}\bar{\t}_1(\bz-\frac kn \bz_1-(1-\frac kn)\bz_2)|\tau)\nonumber \\
\wzlk=\int_0^\tau dz \, \, \t_1(z-z_1|\tau)^{-(1-\frac kn)}\t_1(z-z_2|\tau)^{-\frac kn}\t_1(z-(1-\frac kn)z_1-\frac kn z_2)|\tau) \nonumber \\
\wzzk=\int_0^{\bar{\tau}} d \bz \, \,\bar{\t}_1(\bz-\bz_1|\tau)^{-\frac kn}\bar{\t}_1(\bz-\bz_2|\tau)^{-(1-\frac kn)}\bar{\t}_1(\bz-\frac kn \bz_1-(1-\frac kn)\bz_2)|\tau)\nonumber \\
\eea
and 
\bea
W_{1}^{1*}=W_1^1=W_1^2, \quad W_2^{1*}=-W_2^{1}=W_2^2\,.
\eea

\section*{Aknowledgements}

This work is an application of the very extensive work done in collaboration with S. Naculich on aspects of gauged WZW models, as well as related work in collaboration with D. Karabali, N. Redlich, and K. Tsokos. We thank these colleagues for their contributions to our understanding of this subject. We also thank Matt Headrick and Albion Lawrence for reading the manuscript.

We are grateful to Cesar Ag\'on and Isaac Cohen for their invaluable assistance in preparing this manuscript.

The research of H. J. Schnitzer is supported in part by DOE by grant  DE-SC0009987

\newpage
\bibliographystyle{utphys}

\bibliography{emirefs}

\providecommand{\href}[2]{#2}\begingroup\raggedright\begin{thebibliography}{10}

\bibitem{Chen:2015cna}
B.~Chen and J.-q. Wu, ``{Rényi entropy of a free compact boson on a torus},''
  \href{http://dx.doi.org/10.1103/PhysRevD.91.105013}{{\em Phys. Rev.} {\bf
  D91} (2015) no.~10, 105013},
\href{http://arxiv.org/abs/1501.00373}{{\tt arXiv:1501.00373 [hep-th]}}.

\bibitem{Atick:1987kd}
J.~J. Atick, L.~J. Dixon, P.~A. Griffin, and D.~Nemeschansky, ``{Multiloop
  Twist Field Correlation Functions for $Z(N$) Orbifolds},''
\href{http://dx.doi.org/10.1016/0550-3213(88)90302-1}{{\em Nucl. Phys.} {\bf
  B298} (1988)  1--35}.

\bibitem{Lokhande:2015zma}
S.~F. Lokhande and S.~Mukhi, ``{Modular invariance and entanglement entropy},''
  \href{http://dx.doi.org/10.1007/JHEP06(2015)106}{{\em JHEP} {\bf 06} (2015)
  106},
\href{http://arxiv.org/abs/1504.01921}{{\tt arXiv:1504.01921 [hep-th]}}.

\bibitem{Calabrese:2009ez}
P.~Calabrese, J.~Cardy, and E.~Tonni, ``{Entanglement entropy of two disjoint
  intervals in conformal field theory},''
  \href{http://dx.doi.org/10.1088/1742-5468/2009/11/P11001}{{\em J. Stat.
  Mech.} {\bf 0911} (2009)  P11001},
\href{http://arxiv.org/abs/0905.2069}{{\tt arXiv:0905.2069 [hep-th]}}.

\bibitem{Calabrese:2010he}
P.~Calabrese, J.~Cardy, and E.~Tonni, ``{Entanglement entropy of two disjoint
  intervals in conformal field theory II},''
  \href{http://dx.doi.org/10.1088/1742-5468/2011/01/P01021}{{\em J. Stat.
  Mech.} {\bf 1101} (2011)  P01021},
\href{http://arxiv.org/abs/1011.5482}{{\tt arXiv:1011.5482 [hep-th]}}.

\bibitem{Chen:2014ehg}
B.~Chen and J.-q. Wu, ``{Universal relation between thermal entropy and
  entanglement entropy in conformal field theories},''
  \href{http://dx.doi.org/10.1103/PhysRevD.91.086012}{{\em Phys. Rev.} {\bf
  D91} (2015) no.~8, 086012},
\href{http://arxiv.org/abs/1412.0761}{{\tt arXiv:1412.0761 [hep-th]}}.

\bibitem{Cardy:2014jwa}
J.~Cardy and C.~P. Herzog, ``{Universal Thermal Corrections to Single Interval
  Entanglement Entropy for Two Dimensional Conformal Field Theories},''
  \href{http://dx.doi.org/10.1103/PhysRevLett.112.171603}{{\em Phys. Rev.
  Lett.} {\bf 112} (2014) no.~17, 171603},
\href{http://arxiv.org/abs/1403.0578}{{\tt arXiv:1403.0578 [hep-th]}}.

\bibitem{Chen:2014hta}
B.~Chen and J.-q. Wu, ``{Large Interval Limit of R\'enyi Entropy At High
  Temperature},''
\href{http://arxiv.org/abs/1412.0763}{{\tt arXiv:1412.0763 [hep-th]}}.

\bibitem{Datta:2013hba}
S.~Datta and J.~R. David, ``{Rényi entropies of free bosons on the torus and
  holography},'' \href{http://dx.doi.org/10.1007/JHEP04(2014)081}{{\em JHEP}
  {\bf 04} (2014)  081},
\href{http://arxiv.org/abs/1311.1218}{{\tt arXiv:1311.1218 [hep-th]}}.

\bibitem{Azeyanagi:2007bj}
T.~Azeyanagi, T.~Nishioka, and T.~Takayanagi, ``{Near Extremal Black Hole
  Entropy as Entanglement Entropy via AdS(2)/CFT(1)},''
  \href{http://dx.doi.org/10.1103/PhysRevD.77.064005}{{\em Phys. Rev.} {\bf
  D77} (2008)  064005},
\href{http://arxiv.org/abs/0710.2956}{{\tt arXiv:0710.2956 [hep-th]}}.

\bibitem{Chen:2013kpa}
B.~Chen and J.-J. Zhang, ``{On short interval expansion of Rényi entropy},''
  \href{http://dx.doi.org/10.1007/JHEP11(2013)164}{{\em JHEP} {\bf 11} (2013)
  164},
\href{http://arxiv.org/abs/1309.5453}{{\tt arXiv:1309.5453 [hep-th]}}.

\bibitem{Chen:2015uia}
B.~Chen, J.-q. Wu, and Z.-c. Zheng, ``{Holographic Rényi entropy of single
  interval on Torus: With W symmetry},''
  \href{http://dx.doi.org/10.1103/PhysRevD.92.066002}{{\em Phys. Rev.} {\bf
  D92} (2015) no.~6, 066002},
\href{http://arxiv.org/abs/1507.00183}{{\tt arXiv:1507.00183 [hep-th]}}.

\bibitem{Chen:2015usa}
B.~Chen, W.-Z. Guo, S.~He, and J.-q. Wu, ``{Entanglement Entropy for Descendent
  Local Operators in 2D CFTs},''
\href{http://arxiv.org/abs/1507.01157}{{\tt arXiv:1507.01157 [hep-th]}}.

\bibitem{Caputa:2015tua}
P.~Caputa and A.~Veliz-Osorio, ``{Entanglement constant for conformal
  families},'' \href{http://dx.doi.org/10.1103/PhysRevD.92.065010}{{\em Phys.
  Rev.} {\bf D92} (2015) no.~6, 065010},
\href{http://arxiv.org/abs/1507.00582}{{\tt arXiv:1507.00582 [hep-th]}}.

\bibitem{Chen:2015kua}
B.~Chen and J.-q. Wu, ``{Holographic Calculation for Large Interval R\'enyi
  Entropy at High Temperature},''
\href{http://arxiv.org/abs/1506.03206}{{\tt arXiv:1506.03206 [hep-th]}}.

\bibitem{Chen:2013dxa}
B.~Chen, J.~Long, and J.-j. Zhang, ``{Holographic Rényi entropy for CFT with W
  symmetry},'' \href{http://dx.doi.org/10.1007/JHEP04(2014)041}{{\em JHEP} {\bf
  04} (2014)  041},
\href{http://arxiv.org/abs/1312.5510}{{\tt arXiv:1312.5510 [hep-th]}}.

\bibitem{Headrick:2010zt}
M.~Headrick, ``{Entanglement Renyi entropies in holographic theories},''
  \href{http://dx.doi.org/10.1103/PhysRevD.82.126010}{{\em Phys. Rev.} {\bf
  D82} (2010)  126010},
\href{http://arxiv.org/abs/1006.0047}{{\tt arXiv:1006.0047 [hep-th]}}.

\bibitem{Headrick:2012fk}
M.~Headrick, A.~Lawrence, and M.~Roberts, ``{Bose-Fermi duality and
  entanglement entropies},''
  \href{http://dx.doi.org/10.1088/1742-5468/2013/02/P02022}{{\em J. Stat.
  Mech.} {\bf 1302} (2013)  P02022},
\href{http://arxiv.org/abs/1209.2428}{{\tt arXiv:1209.2428 [hep-th]}}.

\bibitem{Calabrese:2004eu}
P.~Calabrese and J.~L. Cardy, ``{Entanglement entropy and quantum field
  theory},'' \href{http://dx.doi.org/10.1088/1742-5468/2004/06/P06002}{{\em J.
  Stat. Mech.} {\bf 0406} (2004)  P06002},
\href{http://arxiv.org/abs/hep-th/0405152}{{\tt arXiv:hep-th/0405152
  [hep-th]}}.

\bibitem{Calabrese:2009qy}
P.~Calabrese and J.~Cardy, ``{Entanglement entropy and conformal field
  theory},'' \href{http://dx.doi.org/10.1088/1751-8113/42/50/504005}{{\em J.
  Phys.} {\bf A42} (2009)  504005},
\href{http://arxiv.org/abs/0905.4013}{{\tt arXiv:0905.4013
  [cond-mat.stat-mech]}}.

\bibitem{Redlich:1986rp}
A.~N. Redlich, H.~J. Schnitzer, and K.~Tsokos, ``{Bose-fermi Equivalence on the
  Two-dimensional Torus for Simply Laced Groups},''
\href{http://dx.doi.org/10.1016/0550-3213(87)90386-5}{{\em Nucl. Phys.} {\bf
  B289} (1987)  397}.

\bibitem{Schnitzer:1986fi}
H.~J. Schnitzer and K.~Tsokos, ``{Partition Functions and Fermi-bose
  Equivalence for Simple Laced Groups on Compact Riemann Surfaces},''
\href{http://dx.doi.org/10.1016/0550-3213(87)90480-9}{{\em Nucl. Phys.} {\bf
  B291} (1987)  429}.

\bibitem{Hartman:2013mia}
T.~Hartman, ``{Entanglement Entropy at Large Central Charge},''
\href{http://arxiv.org/abs/1303.6955}{{\tt arXiv:1303.6955 [hep-th]}}.

\bibitem{Cardy:2007mb}
J.~L. Cardy, O.~A. Castro-Alvaredo, and B.~Doyon, ``{Form factors of
  branch-point twist fields in quantum integrable models and entanglement
  entropy},'' \href{http://dx.doi.org/10.1007/s10955-007-9422-x}{{\em J.
  Statist. Phys.} {\bf 130} (2008)  129--168},
\href{http://arxiv.org/abs/0706.3384}{{\tt arXiv:0706.3384 [hep-th]}}.

\bibitem{DeNobili:2015dla}
C.~De~Nobili, A.~Coser, and E.~Tonni, ``{Entanglement entropy and negativity of
  disjoint intervals in CFT: Some numerical extrapolations},''
  \href{http://dx.doi.org/10.1088/1742-5468/2015/06/P06021}{{\em J. Stat.
  Mech.} {\bf 1506} (2015) no.~6, P06021},
\href{http://arxiv.org/abs/1501.04311}{{\tt arXiv:1501.04311
  [cond-mat.stat-mech]}}.

\bibitem{Casini:2009sr}
H.~Casini and M.~Huerta, ``{Entanglement entropy in free quantum field
  theory},'' \href{http://dx.doi.org/10.1088/1751-8113/42/50/504007}{{\em J.
  Phys.} {\bf A42} (2009)  504007},
\href{http://arxiv.org/abs/0905.2562}{{\tt arXiv:0905.2562 [hep-th]}}.

\bibitem{Casini:2008wt}
H.~Casini and M.~Huerta, ``{Remarks on the entanglement entropy for
  disconnected regions},''
  \href{http://dx.doi.org/10.1088/1126-6708/2009/03/048}{{\em JHEP} {\bf 03}
  (2009)  048},
\href{http://arxiv.org/abs/0812.1773}{{\tt arXiv:0812.1773 [hep-th]}}.

\bibitem{Casini:2012ei}
H.~Casini and M.~Huerta, ``{On the RG running of the entanglement entropy of a
  circle},'' \href{http://dx.doi.org/10.1103/PhysRevD.85.125016}{{\em Phys.
  Rev.} {\bf D85} (2012)  125016},
\href{http://arxiv.org/abs/1202.5650}{{\tt arXiv:1202.5650 [hep-th]}}.

\bibitem{Liu:2015iia}
F.~Liu and X.~Liu, ``{Two intervals R\'enyi entanglement entropy of compact
  free boson on torus},''
\href{http://arxiv.org/abs/1509.08986}{{\tt arXiv:1509.08986 [hep-th]}}.

\bibitem{Naculich:1989ii}
S.~G. Naculich and H.~J. Schnitzer, ``{Constructive Methods for Higher Genus
  Correlation Functions of Level One Simply Laced {WZW} Models},''
\href{http://dx.doi.org/10.1016/0550-3213(90)90003-V}{{\em Nucl. Phys.} {\bf
  B332} (1990)  583}.

\bibitem{Headrick:2015gba}
M.~Headrick, A.~Maloney, E.~Perlmutter, and I.~G. Zadeh, ``{Rényi entropies,
  the analytic bootstrap, and 3D quantum gravity at higher genus},''
  \href{http://dx.doi.org/10.1007/JHEP07(2015)059}{{\em JHEP} {\bf 07} (2015)
  059},
\href{http://arxiv.org/abs/1503.07111}{{\tt arXiv:1503.07111 [hep-th]}}.

\bibitem{Gukov:2004id}
S.~Gukov, E.~Martinec, G.~W. Moore, and A.~Strominger, ``{Chern-Simons gauge
  theory and the AdS(3) / CFT(2) correspondence},''
\href{http://arxiv.org/abs/hep-th/0403225}{{\tt arXiv:hep-th/0403225
  [hep-th]}}.

\bibitem{Schnitzer:2015gpa}
H.~J. Schnitzer, ``{Left-Right Entanglement Entropy, D-Branes, and Level-rank
  duality},''
\href{http://arxiv.org/abs/1505.07070}{{\tt arXiv:1505.07070 [hep-th]}}.

\end{thebibliography}\endgroup

\end{document}